\pdfoutput=1
\documentclass{article}
\usepackage{booktabs}
\usepackage{multirow}
\usepackage{natbib}
\usepackage[margin=0.5in]{geometry}

\usepackage{amsmath}
\usepackage{amsthm}

\usepackage{booktabs}
\PassOptionsToPackage{round}{package/natbib}
\usepackage{graphicx}
 \usepackage{siunitx}
\usepackage{booktabs} 

\begin{document}


\title{Boost-RS: Boosted Embeddings for Recommender Systems and its Application to Enzyme-Substrate Interaction Prediction}
\author{Xinmeng Li, Li-Ping Liu, and Soha Hassoun \\
Department of Computer Science, Tufts University, Medford 02155, USA
}




\maketitle

\abstract{\textbf{Motivation:}
Despite experimental and curation efforts, the extent of enzyme promiscuity on substrates continues to be largely unexplored and under documented. 
Recommender systems (RS), which are currently unexplored for the enzyme-substrate interaction prediction problem, can be utilized to provide enzyme recommendations for substrates, and vice versa. 
The performance of Collaborative-Filtering (CF) recommender systems however hinges on the quality of embedding vectors of users and items (enzymes and substrates in our case). Importantly, enhancing CF embeddings with heterogeneous auxiliary data, specially relational data (e.g., hierarchical, pairwise, or groupings), remains a challenge. 
\\
\textbf{Results:} 
We propose an innovative general RS framework, termed Boost-RS, that enhances RS performance by  “boosting”
embedding vectors through auxiliary data.  Specifically, Boost-RS is trained and dynamically tuned on multiple relevant auxiliary learning tasks
Boost-RS utilizes contrastive learning tasks to exploit relational data.
 To show the efficacy of Boost-RS for the enzyme-substrate prediction interaction problem,
we apply the Boost-RS framework to several baseline CF models. We show that each of our auxiliary tasks boosts learning of the embedding vectors, and that  contrastive learning using Boost-RS outperforms attribute concatenation and multi-label learning. We also show that Boost-RS outperforms similarity-based models. Ablation studies and visualization of
learned representations highlight the importance of using contrastive learning on some of the auxiliary data in boosting
the embedding vectors.\\
\textbf{Availability and implementation:} A Python implementation for Boost-RS is provided at\\https://github.com/HassounLab/Boost-RS\\
\textbf{Contact:} Liping.liu@tufts.edu and Soha.Hassoun@tufts.edu\\

}

\section{Introduction}
 Understanding the rich functionality of enzymes is fundamental in advancing biochemistry, molecular and synthetic biology and many other application domains. Enzymes were assumed \textit{specific}, catalyzing a specific substrate; however, there is now wide consensus that enzymes are \textit{promiscuous}, catalyzing many substrates, including substrates that the enzymes did not evolve to catalyze \citep{tawfik2010enzyme}. 
 Our ability to analyze this inherent promiscuity has proved instrumental in guiding the direct evolution of novel proteins \citep{romero2009exploring}, elucidating metabolism in natural and engineered organisms \citep{Porokhin2021Analysis}, and creating novel synthesis pathway to produce valuable therapeutics and commodity molecules \citep{bowie2020synthetic}. Despite progress in protein function annotation and modeling protein-ligand interactions (mostly focused on drug-ligand interactions), and  manual and automated curation efforts
, there remains large gaps in our knowledge of enzyme capabilities. Computational tools that predict enzyme promiscuity on molecules can augment existing knowledge, guide biological and biomedical applications and reduce costly experimental efforts. 
 
Computational approaches for predicting enzyme-substrate interactions target different applications. 
Physics-based models, including molecular docking and molecular dynamic simulations, attempt to identify the most favorable binding mode of a ligand with a given target protein 
These methods require 3D models of both protein and molecule, and require significant compute time, making these methods suitable for detailed analysis of a small number of interactions.  
Rule-based methods predict site of metabolism or products of enzymatic transformations on a query molecule. Most such methods, however, utilize hand-curated  biotransformation rules (e.g., \citep{ridder2008sygma}), or applicable to only specific enzymes (e.g., \citep{tyzack2019computational}), thus limiting their general applicability. 
Machine-Learning (ML) approaches have taken advantage of available enzymatic data and solve many important questions such the likelihood of enzymatic transformations between a compound pair, e.g. Support Vector Machines \citep{kotera2013kcf}, graph embedding \citep{jiang2021learning}, identifying enzyme commission numbers that act on molecules, e.g. using hierarchical classification of enzymes on molecules \citep{visani2020enzyme}, and predicting the likelihood of a sequence catalyzing a reaction or  quantifying the affinity of sequences on substrates using Gaussian processes \citep{mellor2016semisupervised}. Once trained, ML models provide quick evaluation and are suited for many bioengineering and biological applications that require the exploration of the vast interaction space. 

To expand the use of ML in predicting enzyme-substrate interactions, we investigate the use of recommender systems  to recommend enzymes that are likely to act on specific substrates, and/or compounds that are suited as substrates for an enzyme. Recommender systems (RS) are heavily utilized in industrial applications. For example, more than 50\% of all AI training cycles at Facebook are devoted to training deep learning recommendation models \citep{acun2021understanding}. RS, however, were not used prior for predicting enzyme-substrate interactions.  Previously, RS were used for predicting protein-drug interactions  \citep{bagherian2021machine}.  Many such techniques employ collaborative filtering (CF) in the form of matrix factorization (MF), e.g., Multiple Similarities Collaborative Matrix Factorization (MSCMF) \citep {zheng2013collaborative}, Probabilistic Matrix Factorization (PMF) \citep{mnih2008probabilistic}, Neighborhood Regularized Logistic Matrix Factorization (NRLMF) \citep{liu2016neighborhood}. 

As the performance of CF hinges on learned embeddings of the users and items, prior RS techniques aimed to utilize auxiliary (side) data to learn improved embeddings.  
Many techniques enhance CF by using similarities among proteins and among drugs, e.g., MSCMF \citep {zheng2013collaborative} or neighborhood regularization, e.g., NRLMF \citep{liu2016neighborhood}, and REMAP \citep{lim2016large}, with the goal of minimizing  distances between a protein (or a drug) and its nearest neighbors in the latent space.  Other RS aim to integrate auxiliary data by fusing  knowledge graphs (e.g., \citep{wang2021multitask}), or integrating multi-source data (e.g.,  \citep{zhu2017broad}, \citep{gao2018recommendation}). In practice, auxiliary data is complex and often exhibits multiple relational aspects: item labels may be hierarchical, and users may share a group label (zip code, building address, or profession). The common practice to concatenate auxiliary data with the learned embedding does not necessarily maximally exploit the relational aspect of the data. A general methodology for computing enhanced embeddings based on relational data and other complex heterogeneous auxiliary data therefore remains a challenge.

We present in this paper a novel technique, Boost-RS, for enhancing the performance of RS by "boosting" the embedding vectors through auxiliary learning tasks.  Boost-RS integrates the primary CF task with boosting tasks that aim to upgrade the embedding vectors based on available heterogeneous auxiliary data. The integration of user and item attributes  addresses the interaction matrix sparsity issue and has already shown  RS performance improvements \citep{sun2019research, bagherian2021machine}.  To minimize negative transfer from the auxiliary tasks to the main task, the CF and the boosting tasks are dynamically weighted \citep{liu2019loss}.  Each auxiliary task is designed to maximally utilize the available auxiliary data. Importantly, to learn from relational data, Boost-RS employs contrastive learning, which contrasts positive and negative samples to learn discriminative representations in a self-supervised manner. 
Contrastive learning is applied through triplet loss \citep{weinberger2009distance}, where the model is trained to produce representations such that, for a given anchor example, a positive example is closer to the anchor than a negative example.  Prior RS contrastive work \citep{liu2021contrastive} used perturbation to user (or item) preferences and contrasted the perturbed views to maximize learning mutual information between the two views. 

We demonstrate Boost-RS’s effectiveness by applying it to the enzyme-substrate interaction prediction problem. The CF primary task of interaction prediction is accompanied by   multi tasks that learn from heterogeneous auxiliary data. For  enzymes, we exploit the Enzyme Commission  (EC) hierarchical relationships and the enzyme functional orthologs. For  substrates, we utilize the molecular fingerprints and substrate-substrate biotransformation relationships due to 
 functionally similar enzymes.  We use contrastive learning on the enzyme functional orthologs and the biotransformation relationships, and formulate a hierarchical loss on the EC relationships. Our main contributions are:

\begin{itemize}
    \item Creating a flexible and generalizable framework, Boost-RS, that enriches the embedding vectors for CF-based recommender systems using individual and relational heterogeneous auxiliary data.

    \item Showing that using  contrastive learning on relational data may outperform other techniques such as  multi-labels learning and  concatenation with learned embeddings.
    
    \item Demonstrating the generality of the Boost-RS framework by  showcasing its applicability to three recent neural network baseline CFs: Deep Matrix Factorization (DMF) \citep{xue2017deep}, Neural Graph Collaborative Filtering (NGCF) \citep{wang2019neural}, and Neural Matrix Factorization (NMF) \citep{he2017neural}.

     \item  Showing that Boost-RS outperforms  state-of-the-art similarity-based  Graph Regularized Generalized Matrix Factorization (GRGMF) recommender systems \citep{zhang2020graph}.
      
\end{itemize}

\section{Methods}
\subsection{Dataset}

\begin{figure*}
\centerline{\includegraphics[bb=0 0 5309 2411, width=0.95\textwidth]{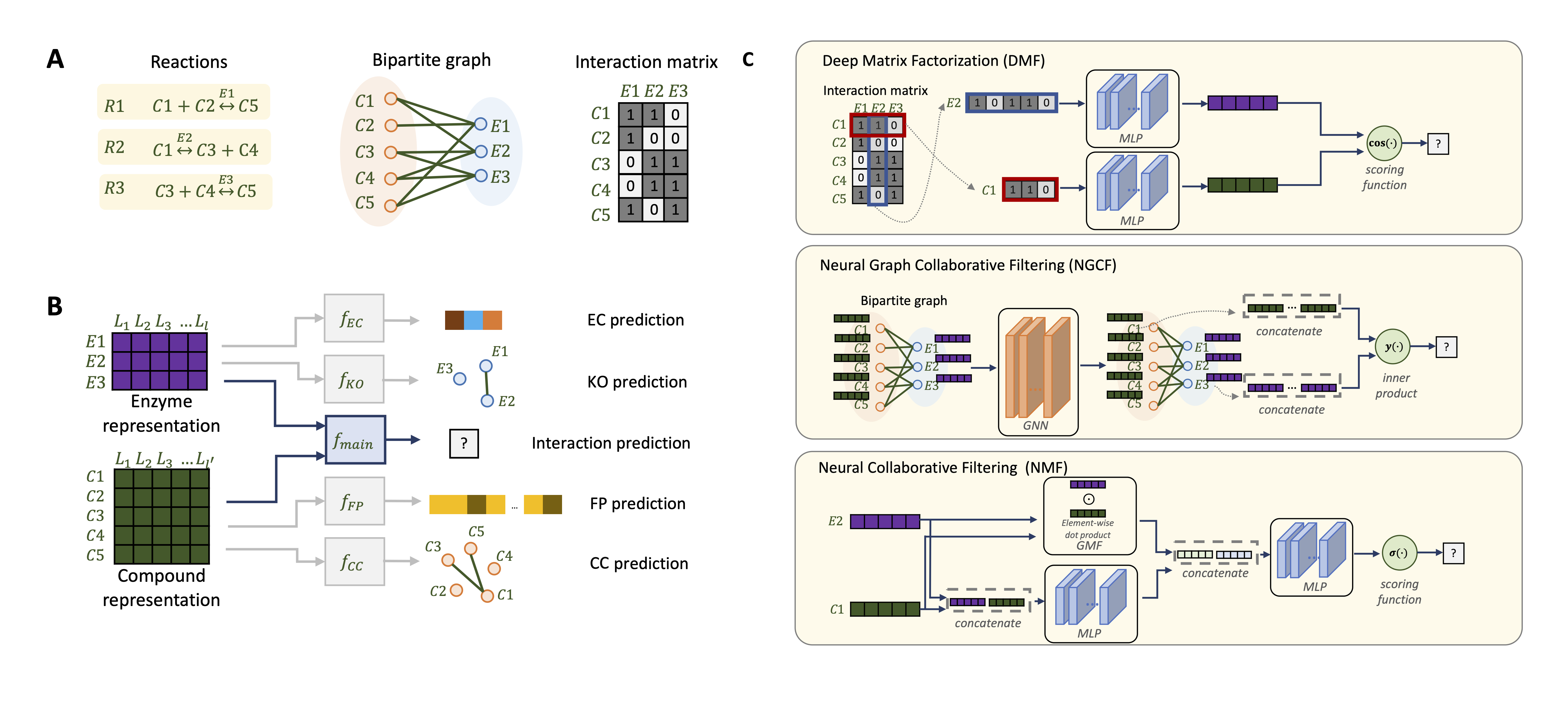}}
\caption{\textbf{Boost-RS framework for enzyme-substrate recommendation prediction.}  A) Interaction matrix construction from enzymatic reactions. e.g., for $E1$, three positive interactions are  added to the matrix. B) The Boost-RS framework that integrates the main task of interaction prediction with related auxiliary tasks. C) Collaborative filtering models used as baselines.} \label{fig:model}
\end{figure*}

We apply Boost-RS to the enzyme-substrate interaction prediction task to recommend substrates to enzymes that are most likely to interact and vice versa.  Our dataset is culled from biochemical reactions in the KEGG database \citep{kanehisa2000kegg}. 
As most biochemical reactions are reversible, no distinction is made between substrates and products, and hence interacting molecules are referred to as compounds or substrates interchangeably. 
Reactions form a bipartite graph that can be captured as a binary interaction matrix between enzymes and substrates  (Fig. \ref{fig:model}A), where each row represents a compound, and each column represents an enzyme. 
A matrix entry is set to $1$ if the compound and enzyme participate the same reaction, therefore representing a positive interaction instance. An entry is set to $0$ in case there is no catalogued enzyme-substrate interaction. 
Compounds that are common to many enzymatic reactions, including cofactors such as ATP and NADH and metals, are excluded from the matrix. To ensure valid data splits, enzymes or compounds with a single entry in the interaction matrix are also excluded. In total, 17,627 enzyme-compound interactions are collected. They involve 4,768 enzymes and 6,397 compounds. The interaction matrix is therefore sparse with 0.06\% positive entries.

\subsection{Auxiliary Data}
Four attributes are collected or derived from the KEGG database and are used for training auxiliary tasks:

\textbf{Enzyme Commission (EC) numbers. }Each enzyme is associated with an EC number \citep{webb1992enzyme} that comprises four numbers, separated by dots, starting with a number that broadly represents the enzyme class, then the sub-class, sub-subclass, and a final number that reflects the specificity of the enzyme towards a small group of substrates. As the EC numbers are hierarchical, the auxiliary learning task of EC prediction can be formulated as such. The EC numbers have 7, 94, and 164 distinct labels in the first three respective fields of EC. We only consider the first three fields of EC, as the fourth index typically denotes specific substrates and cofactors.
    
\textbf{Functional Orthologs (KO) numbers.} Another enzyme attribute is its KEGG functional orthology (KO) number \citep{kanehisa2016kegg}. A particular KO designation, the letter K followed by 5 numerical digits, is assigned to a group of genes sharing similar  functionality.  KO numbers can therefore be considered as a "group" attribute. We utilize 5575 sets of KO designations. 

\textbf{Molecular fingerprints (FP). } Based on descriptions for molecules in the KEGG database, molecular attributes in the form of MACCS fingerprints (FP) \citep{durant2002reoptimization} are calculated using RDKIT. 


\textbf{Compound-compound biotransformations (CC).} The KEGG database provides  biotransformation patterns, designated as RClasses, shared by multiple substrate-product, or compound-compound (CC), pairs.
CC pairs under the same RClass are transformed by enzymes with similar functionality (e.g., hydroxylation or methylation). 
CC relationships give rise to a compound-centric graph, akin to a social network. 
This graph is \textit{not} a similarity network as CC pairs are not necessarily similar: a compound may undergo significant molecular changes under some enzymatic transformations (e.g., transferases, ligases). 


\subsection{The Boost-RS model}
\subsubsection{Main task: interaction prediction}
The main task of the Boost-RS framework (Fig. \ref{fig:model}B) is ``recommending'' compounds to enzymes (or enzymes to compound). Per the interaction matrix, an entry $y_{ij}$ is $1$ for the positive interaction set, $P$, and $y_{ij}$ is $0$ for the negative interaction set, $N$. As in a standard recommendation task, the main task function, $f_{main}(\cdot, \cdot)$, predicts the probability $\hat{y}_{ij}$ of interaction using learned enzyme and compound representations. We denote the representations for compound $i$ and enzyme $j$ as  $\mathbf{v}_i$ and $\mathbf{u}_j$, respectively. The probability of interaction, $\hat{y}_{ij}$, and task loss, $\mathcal{L}_{main}$, are then defined as:

\begin{gather}
    \hat{y}_{ij} = f_{main}(\mathbf{v}_i, \mathbf{u}_j)\\
    \mathcal{L}_{main} =  \sum_{i,j} BCE(\hat{y}_{ij}, y_{ij})
\end{gather}


In the simplest form, $f_{main}(\mathbf{v}_i, \mathbf{u}_j) = \mathbf{v}_i^\top \mathbf{u}_j$. 
In more complex forms,  $f_{main}(\mathbf{v}_i, \mathbf{u}_j)$ is calculated using neural networks.
The parameters for $f_{main}$ and the representations of enzymes and compounds are updated by minimizing the task loss.
The specifics of this loss are dependent on the underlying base CF model used as a RS (see Section \ref{sec:baselineModels}). 
The embedding vectors, $\mathbf{v}_i$ and $\mathbf{u}_j$, are critically important for the recommendation task.
To boost  performance, we inject task-relevant auxiliary data for compounds and enzymes into the embedding vectors via multi-task learning.

\subsubsection{Auxiliary tasks}
Each auxiliary task calculates attribute probabilities with an auxiliary task function and updates  representations and parameters based on the task loss. 
To address the differing characteristic in relational attributes, auxiliary losses are defined on individual, hierarchical, group, and pairwise attributes. While we describe the details relevant to the specific substrate and enzyme attributes, the framework easily accommodates other attributes with individual and relational attributes.

Using FP as an individual attribute, we denote the function for FP prediction as  $f_{FP}$. 
As each compound fingerprint is a binary vector, we use BCE to evaluate the prediction accuracy for each vector entry. 
The FP prediction function and the FP task loss are defined as:
\begin{gather}
    \hat{y}^{FP}_i = f_{FP}\left(\mathbf{v}_i\right),\\
    \mathcal{L}_{FP} = \frac{1}{|C|} * \frac{1}{l_{FP}} \sum_{i \in C} 
      BCE(\hat{y}_{i}^{FP}, y_{i}^{FP}),
\end{gather}
where $C$ is the set of compounds, and $l_{FP}$ is the length of the fingerprint vector.

For EC prediction, we capitalize on the EC's hierarchical structure, and denote the function for  EC prediction as $f_{EC}$.
The loss for each enzyme is based on the cross entropy loss 
on each of its field set, $FS_{1}$, $FS_{2}$, $FS_{3}$ for the first, the first two, and the first three fields of the EC number, respectively.
The prediction function and the cumulative task loss for the EC attribute is therefore: 
\begin{gather}
        \hat{y}^{EC}_{j,FS_{k}} = f_{EC}\left(\mathbf{u}_j\right),\\
    \mathcal{L}_{EC} = \frac{1}{|E|} \sum_{j \in E} \sum_{k=1}^{3} w_k \cdot CE(\hat{y}^{EC}_{j,FS_{k}}, y^{EC}_{j,FS_{k}}),
\end{gather}
where $E$ is the set of enzymes, and $k$ is indicating which field set, $FS$, in EC are we calculating the cross entropy loss on its distinct labels. And $w_k$ is the weight for the $FS_{k}$ field set in EC, and, after some tuning, is set to $\frac{1}{3}$, therefore equalizing each field's contribution to the loss.

When using group attributes and pairwise relationship for the auxiliary tasks, we denote functions for KO prediction and CC prediction as $f_{KO}$ and $f_{CC}$, respectively. We utilize triplet loss, with the intention of pulling the representation of samples in the same set closer together in the embedding space and pushing away the representation of a sample outside the set. A function, $d(\cdot)$, measures the distance between a pair of representations. The CC loss function is defined as:

\begin{align}
    \mathcal{L}_{CC} = \frac{1}{|C|} \sum_{i,P_i, N_i\in C} \max(0, d\left(\mathbf{v}_i, \mathbf{v}_{P_i}\right) - d\left(\mathbf{v}_i, \mathbf{v}_{N_i}\right) + \gamma),
\end{align}
where $i$ an anchor compound, $P_i$ is a compound in the set of compounds has CC pairwise relationship with compound $i$,  $N_i$ is a compound in a set of compounds that does not have an CC relationship with compound $i$, and $\gamma$ is a positive margin between the positive $P_i$ and negative $N_i$ samples.

The KO loss function is defined similarly to the CC loss function, except that the anchor, positive and negative samples are derived from the KO group relationships defined on the enzymes. The contrastive loss on CC reflect a pairwise relationship within the compound-centric relational graph, while the loss on KO reflects the group label distinction.


\subsubsection{Dynamic training}
The Boost-RS loss, $\mathcal{L_\text{Boost-RS}}$, is the weighted sum of losses of the main task and auxiliary tasks: 
\begin{align}
    \mathcal{L_\text{Boost-RS}} = \alpha_{main} \, \mathcal{L}_{main} + \alpha_{aux} \, \mathcal{L}_{aux},
\end{align}
where $L_{aux}$ is the additive losses across the auxiliary tasks, $\alpha_{main}$ and $\alpha_{aux}$ are weights for the main task and auxiliary tasks, respectively. 

The weights of task losses directly influence the performance of interaction prediction. We employ a dynamic strategy to balance the weights among the main task and auxiliary tasks. The abridged linear schedule \citep{belharbi2016deep} 
emphasizes auxiliary tasks in the early training epochs and shifts the focus to the main task in later epochs. We assign weights for the main and auxiliary tasks losses as follows:
\begin{gather}
     \alpha_{main} = \min(\frac{t}{T}, 1) ; \alpha_{aux} = \max(1 - \alpha_{main}, 0),
\end{gather}
where $t$ is the current training iteration, and $T$ is time point where the focus shifts completely to  the main task. 

\subsubsection{Deep-learning baseline recommender systems}
\label{sec:baselineModels}
We use three neural-network  recommender systems  (Fig. \ref{fig:model}C) for the interaction prediction task: DMF \citep{xue2017deep}, NGCF \citep{wang2019neural}, and NMF \citep{he2017neural}. Each recommender system has its own characteristic and may be better suited for some applications. We apply Boost-RS to each of these models.

The inputs to DMF are the rows and columns of the interaction matrix. Two separate Multi-layer Perceptron (MLP) networks are trained through $f_{main}$ to learn  compound and enzyme representations. The similarity of the enzyme and compound  representations are computed using   cosine similarity and outputed as the  probability of the interaction, $\hat{y}$. Normalized cross entropy loss is used to compute the loss between $y$ and $\hat{y}$. 

To compute $f_{main}$, NGCF first applies graph neural network to the bipartite interaction graph. Graph Neural Network (GNN) are utilized to learn  node representations.  GNNs can account for different order neighbors including first order neighbors, second order neighbors, and so on.  Node representations for enzyme (compound) nodes that are learned for each level of neighbors are then concatenated. The inner product of the enzyme and compound representations is then computed. 
To compute the loss, an additional optimization step (not shown in the figure) favors 
assigning higher predictions for known interactions than for unknown interactions. 

For NMF, $f_{main}$ has GMF and MLP working in parallel and are then followed with a scoring layer. GMF and MLP each learn independent representations. A function $\sigma$  calculates the interaction probability based on the concatenation of the learned representations.

\section{Results}
\subsection{Experimental setup}
We divide positive interactions into training, validation, and test sets at a ratio of 7:2:1. 
During training and validation, negative interactions are randomly sampled from the unknown interactions at a negative sampling ratio, which is a hyperparameter that varies across models.  During testing, all unknown interactions are assumed negative. Positive interactions in the training set are excluded  during sampling.

As the test data set is  imbalanced, where the ratio of positive to the assumed negative interactions is less than 0.01\%, evaluation metrics are selected to reflect the ability of  RS to rank  positive interactions ahead of negative ones.  Average Precision (AP) computes the average precision after each predicted positive interaction in the ranked order list provided by RS. 
AP is utilized for model selection.  
To place lower emphasis on the exact ranking of known interactions, R-Precision computes the  precision after all R positive interactions have been identified in the ranked order list. The overall performance at distinguishing between positive and negative interactions is reported using the area under the receiver operating characteristic curve (AUC).  
We additionally report the Mean AP (MAP) and the R-Precision across the enzymes and the substrates. As each enzyme and substrate had a varied number of positives, we also  report the MAP for the top 3 items (MAP@3) and the precision on the top one item (Precision@1). 

For the three baseline recommender systems, we follow the authors' guidelines on hyperparameter tuning. The range of hyperparameter search is specified as follows. The negative sampling ratio for training set is selected from $\{1, 5, 10, 15, 20, 25, 30\}$. The margin in the triplet loss, $\gamma$, is selected from $\{0.5, 1.0, 1.5\}$. The dimension of the two hidden layers of MLPs of $f_{main}(\cdot), f_{FP}(\cdot), f_{EC}(\cdot)$ predictor is selected from \{128, 256, 512\}. We optimize our models with the Adam optimizer \citep{kingma2015adam} with learning rates selected among $\{10^{-1}, 10^{-2}, 10^{-3}, 10^{-4}\}$. We apply dropout at a rate selected from $\{0.0, 0.3, 0.5\}$ and L2 norm at a weight selected from $\{10^{-2}, 10^{-3}, \ldots, 10^{-6}\}$. For the abridged linear dynamic weighting strategy, we allow a maximum of 3000 iterations, with $T=2000$. During the first 2000 iterations, the model shifts linearly from training the auxiliary tasks to the interaction prediction task. Training is  stopped early if there is no improved MAP on the validation set in 500 consecutive iterations.

\begin{table*}[!t]
\tiny
\caption{Interaction-prediction performance evaluation. The best model (\textbf{Boost-RS}) is based on NMF and utilizes contrastive learning on KO and CC, hierarchical learning on EC, and individual attribute learning on FP. } 
\begin{tabular} {cccccccccccc}

\toprule 
 &  \multicolumn{3}{c}{Overall} & \multicolumn{4}{c}{Enzymes}& \multicolumn{4}{c}{Compounds} \\
\cmidrule(lr){2-4} \cmidrule(lr){5-8} \cmidrule(lr){9-12}
& AP & R-Precision & AUC & MAP & R-Precision & MAP@3 & Precision@1  & MAP & R-Precision & MAP@3 & Precision@1  \\
\midrule
 &  \multicolumn{11}{c}{A. Baselines and their boosted models}\\
\cmidrule(lr){2-12}
DMF  & 0.154 & 0.344 & 0.869 & 0.328 & 0.251 & 0.261 & 0.255  & 0.281 & 0.282 & 0.334 & 0.332   \\
Boost-DMF   & 0.192 & 0.401 & 0.954 & 0.374 & 0.258  & 0.273 & 0.294 & 0.309 & 0.296 & 0.362 & 0.374   \\

NGCF  & 0.169 & 0.328 & 0.810& 0.333 & 0.274 & 0.278 & 0.261  & 0.277 & 0.297 & 0.347 & 0.326   \\
Boost-NGCF  & 0.223 & 0.503 & 0.959 & 0.552 & 0.295 & 0.446& 0.393  & 0.405 & 0.488  & 0.566 & 0.490 \\

NMF  & 0.280 & 0.380 & 0.880 & 0.339 & 0.322 & 0.286 & 0.309  & 0.332 & 0.320 & 0.362 & 0.378   \\
\textbf{Boost-RS} (Boost-NMF) & \textbf{0.488} & \textbf{0.638} & \textbf{0.968} & \textbf{0.595} & \textbf{0.510} & \textbf{0.506} & \textbf{0.545}  & \textbf{0.568} & \textbf{0.546} & \textbf{0.620} & \textbf{0.637}  \\

\cmidrule(lr){1-12}
&  \multicolumn{11}{c}{B. Group data treated as individual attributes and incorporated into RS via either multi-tasking or concatenation}\\
\cmidrule(lr){2-12}

{Boost-RS\_Multi-label}  & 0.404 & 0.492 & 0.936 & 0.419 & 0.411 & 0.353 & 0.421  & 0.452 & 0.395 & 0.441 & 0.490  \\
{NMF-Concat\_Multi-label} & 0.396 & 0.485 & 0.950 & 0.430 & 0.406 & 0.363 & 0.413  & 0.441 & 0.408 & 0.454 & 0.480 \\

\cmidrule(lr){1-12}
 &  \multicolumn{11}{c}{C. Interaction prediction with each auxiliary task using Boost-RS}\\
 \cmidrule(lr){2-12}
Boost-RS(KO) & 0.296 & 0.377 & 0.857 & 0.321 & 0.315 & 0.280 & 0.321  & 0.349 & 0.319 & 0.347 & 0.381  \\
Boost-RS(FP) & 0.309 & 0.402 & 0.914 & 0.370 & 0.333 & 0.307 & 0.327  & 0.350 & 0.340 & 0.388 & 0.395  \\
Boost-RS(EC) & 0.344 & 0.432 & 0.880 & 0.337 & 0.377 & 0.294 & 0.372  & 0.399 & 0.325  & 0.364 & 0.438 \\

Boost-RS(CC) & 0.419 & 0.548 & 0.936 & 0.527 & 0.447 & 0.447 & 0.467  & 0.492 & 0.487 & 0.553 & 0.546  \\

\cmidrule(lr){1-12}
 &  \multicolumn{11}{c}{D. Interaction prediction with each auxiliary data using NMF-Concat\_Multi-label}\\
 \cmidrule(lr){2-12}
NMF-Concat(KO) & 0.285 & 0.386 & 0.872 & 0.346 & 0.327 & 0.296 & 0.319  & 0.343 & 0.337 & 0.375 & 0.384 \\
NMF-Concat(FP) & 0.287 & 0.386 & 0.870 & 0.338 & 0.329  & 0.286 & 0.318 & 0.344 & 0.324 & 0.368 & 0.386 \\
NMF-Concat(EC) & 0.292 & 0.390 & 0.879 & 0.339 & 0.333& 0.289 & 0.324  & 0.349 & 0.320 & 0.361 & 0.392 \\
NMF-Concat(CC) &  0.322&  0.408 & 0.868&  0.351 & 0.343 & 0.297 & 0.351&  0.372&  0.342 & 0.380&  0.412\\

\cmidrule(lr){1-12}
 &  \multicolumn{11}{c}{E. Interaction prediction comparing Boost-RS framework against similarity-based method}\\
\cmidrule(lr){2-12}
GRGMF(FP+EC) & 0.189 & 0.407 & 0.946 & 0.362 & 0.282 & 0.266 & 0.293  & 0.307 & 0.281 & 0.361 & 0.387  \\

Boost-RS(FP+EC) & 0.349 & 0.456 & 0.931 & 0.376 & 0.377 & 0.314 & 0.385 & 0.408 & 0.352 & 0.398 & 0.453  \\
\bottomrule
\end{tabular}
\label{Tab:01}

\end{table*}

\subsection{Evaluating Boost-RS on baseline models}
We evaluate the performance gain (Table \ref{Tab:01}A) when implementing Boost-RS for the three baselines Boost-RS  significantly boosts the performance of every baseline across all metrics. 
NMF, which is the best performing baseline, gains  74\%, 68\%, and 10\% on MAP, R-Precision, and AUC, respectively, when combined with Boost-RS. 
For this work, we use Boost-NMF as our  "Boost-RS" model, and use NMF as a baseline model for the rest of the experiments, unless noted otherwise.  Boost-RS utilizes contrastive learning on KO and CC,
hierarchical learning on EC, and individual attribute learning on FP.

\subsection{Multi-label learning via multi-tasking or concatenation}
We create a compound (enzyme) binary multi-label vector for CC (KO), where each entry of the vector indicates the presence or absence of an RClass (KO) designation. The length of the vector is the number of distinct CCs (KOs). For CC, the multi-label vector has 3163 entries, where each compound is involved on average with 1.03 distinct CCs. For KO, the multi-label vector has a length of 5575 entries, where each enzyme has on average 1.37 distinct KO designations. 

We then compare multi-label performance using contrastive loss (Boost-RS\_Multi-label), where we use weighted BCE for the multi-label loss, and using concatenation (NMF-Concat\_Multi-label), where we concatenate the outputs of GMF and MLP with the outputs of MLP layers representing the auxiliary data. The comparisons  (Table \ref{Tab:01}B)  utilize multi-label KO and CC data along with hierarchical EC and FP attributes.
Boost-RS\_Multi-label outperforms NMF-Concat\_Multi-label as  contrastive learning explicitly enforces negative pairs to have distinct representations, while multi-label training does not. 
Further, the sparsity of the KO and CC multi-label vectors (less than 0.1\% of non-zero vector entries) may contribute to the limited improvements using concatenation.
Importantly,  a flexible framework such as Boost-RS allows the judicious selection of the appropriate losses to boost the embeddings (last row of Table \ref{Tab:01}A).

\begin{figure*}[!ht]
\centerline{\includegraphics[bb=0 0 5333 3000, width=0.95\textwidth]{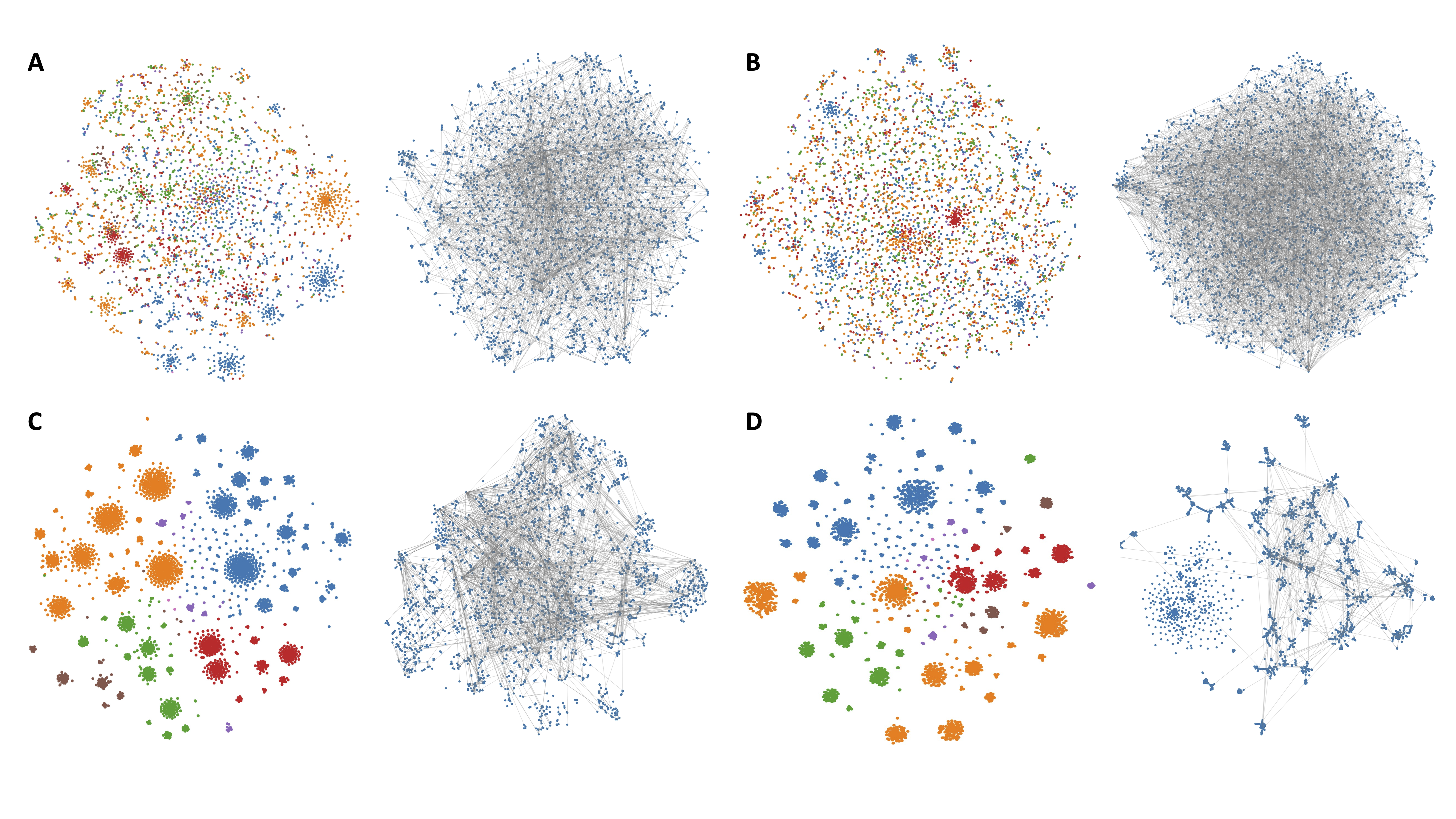}}
\caption{Visualization using t-SNE for learned representation of enzymes and compounds, shown to the left and right of each sub-panel, respectively. A) Baseline NMF. B) Baseline with multi-label KO and CC concatenation (NMF-Concat\_Multi-label). C) Boost-RS with the auxiliary task of learning multi-label KO and CC (Boost-RS\_Multi-label). D) Boost-RS with triplet loss on KO and CC (Boost-RS).}\label{fig:tsne}
\end{figure*}

We use t-SNE \citep{van2008visualizing} to visualize learned enzyme and compound representations (Fig. \ref{fig:tsne}) using the various techniques (NMF, NMF-Concat\_Multi-label, Boost-RS\_Multi-label, and Boost-RS). Enzyme representations are shown to the left of each sub-panel. Compound representations are shown to the right of each sub-panel, where an edge is added between two compound representations if the two compounds are related via a CC. 
For the enzyme representations, each dot is colored with an EC class. Enzymes in sub-panels C and D form the most distinguishable clusters when compared to  sub-panels A and B as both multi-label and contrastive learning perform well on KO (see next section). Across the sub-panels, compound representations initially show no evident pattern (sub-panel A), but 
display more defined clusters with the progression towards sub-panel D. 
There is a node grouping on the left side of the D sub-panel for compounds that lack a CC relation as evident by the absence of any edges connecting these compounds.

\subsection{Contributions of individual auxiliary tasks}
We characterize the contribution of each auxiliary task to the performance of Boost-RS independently of other tasks (Table \ref{Tab:01}C). We also contrast each such contribution against using the same data via concatenation with the baseline NMF model (Table \ref{Tab:01}D).
For both Boost-RS and NMF-Concat, each auxiliary task contributes positively to predicting the overall interactions, indicating that these auxiliary tasks are relevant to the main task and provide additional information beyond what is captured within CF. 
For Boost-RS, 
CC contributes the most (50\% improvement on the overall AP), while while KO, FP, and EC show more modest improvements (6\%, 10\%, 23\% respectively).
For NMF-Concat, CC improves the baseline NMF by 15\% AP, while KO, FP and EC show limited improvements (2\%, 3\%, 4\%, respectively). 
 Boost-RS(CC) improves enzyme and compound MAPs significantly (enzyme MAP by 55\% and compound MAP by 48\%).
These results indicate that each auxiliary data improves the learning of  compound and enzyme representations. 

Boost-RS consistently improves performance on each tasks over NMF-concat, with the exception of the KO prediction task on select metrics other than overall AP, Precision\@1 for enzymes, and MAP for the compounds. 
Despite this performance variation and other experimental evaluations, learning KO using contrastive learning results in the best boosting performance (last row in Table \ref{Tab:01}A) as our model selection is based on AP.
We attribute this varied performance to the characteristics of the KO auxiliary data. For KO, contrastive loss is applied on 779 paired relationships, where there are multiple pairwise relationships involving all pairs of enzymes under the same KO group label. In contrast, for CC, contrastive loss is applied to 7915 paired relationships derived from pairwise compound-compound transformations.

\subsection{Boost-RS vs similarity-based models}
We compare Boost-RS with a recent recommender system, Graph Regularized Generalized Matrix Factorization (GRGMF) \citep{zhang2020graph}. In addition to implementing matrix factorization, GRGMF learns 
latent node representations based on their neighborhood similarity.
GRGMF therefore provides an alternative model for incorporating EC and FP auxiliary data. GRGMF takes as input a pairwise compound-similarity matrix and a pairwise enzyme-similarity matrix, where we use EC numbers and FP to obtain Jaccard similarity scores. CC and KO relationships cannot be readily integrated with GRGMF. We therefore evaluate Boost-RS when using only EC and FP as auxiliary data. 
Boost-RS outperforms GRGMF in most metrics, except for the AUC of GRGMF (Table \ref{Tab:01}E). 
The results show that the Boost-RS framework can effectively capture the auxiliary data.

\section{Conclusion}
Our proposed  framework, Boost-RS,  offers an elegant and generalizable model for boosting learned representations with heterogeneous auxiliary data for CF recommender systems. Dynamically training Boost-RS on multiple tasks allows  evolving their relative weights along the learning epochs. The learning tasks are applicable to various individual and relational attributes. While intended as a  general framework, we here demonstrated the utility of Boost-RS for enzyme-substrate interaction prediction task.  We demonstrated Boost-RS on three CF baseline models.  We identified four auxiliary data (molecular fingerprints, enzyme commission numbers, functional orthologs and  biotransformation relationships), and proved their relevance for enhancing interaction prediction. 
Importantly, we showed the flexibility of Boost-RS framework through multi-task learning allows the integration of various auxiliary data modalities such as individual attributes, group attributes, pairwise relationship.  Replacing relational attributes with their multi-label representations and using them concatenation or multi-task learning, cannot achieve the same performance boost as with Boost-RS. 
We compared Boost-RS with  similarity-based RS models and showed that Boost-RS outperforms GRGMF when utilizing the same data.  Because of its demonstrated advantages, generality and elegance in integrating attributes with CF, the Boost-RS framework may prove beneficial for a diverse set of application.  Further, the use of the trained auxiliary machinery might prove useful in addressing the cold-start problem, common across recommender systems.

\section*{Acknowledgements}

\section*{Funding}
This research is supported by NSF Award 1909536. Li-Ping Liu is supported by NSF Award 1850358 and NSF Award 1908617. The research is also supported by the NIGMS of the National Institutes of Health, Award R01GM132391. The content is solely the responsibility of the authors and does not necessarily represent the official views of the NIH.
\vspace*{-12pt}

\bibliographystyle{abbrvnat}

\bibliography{reference}

\begin{thebibliography}{}

\bibitem[Acun {\em et~al.}(2021)Acun, Murphy, Wang, Nie, Wu, and
  Hazelwood]{acun2021understanding}
Acun, B., Murphy, M., Wang, X., Nie, J., Wu, C.-J., and Hazelwood, K. (2021).
\newblock Understanding training efficiency of deep learning recommendation
  models at scale.
\newblock In {\em 2021 IEEE International Symposium on High-Performance
  Computer Architecture (HPCA)\/}, pages 802--814. IEEE.

\bibitem[Bagherian {\em et~al.}(2021)Bagherian, Sabeti, Wang, Sartor,
  Nikolovska-Coleska, and Najarian]{bagherian2021machine}
Bagherian, M., Sabeti, E., Wang, K., Sartor, M.~A., Nikolovska-Coleska, Z., and
  Najarian, K. (2021).
\newblock Machine learning approaches and databases for prediction of
  drug--target interaction: a survey paper.
\newblock {\em Briefings in bioinformatics\/}, {\bf 22}(1), 247--269.

\bibitem[Belharbi {\em et~al.}(2016)Belharbi, H{\'e}rault, Chatelain, and
  Adam]{belharbi2016deep}
Belharbi, S., H{\'e}rault, R., Chatelain, C., and Adam, S. (2016).
\newblock Deep multi-task learning with evolving weights.
\newblock In {\em ESANN\/}.

\bibitem[Bowie {\em et~al.}(2020)Bowie, Sherkhanov, Korman, Valliere,
  Opgenorth, and Liu]{bowie2020synthetic}
Bowie, J.~U., Sherkhanov, S., Korman, T.~P., Valliere, M.~A., Opgenorth, P.~H.,
  and Liu, H. (2020).
\newblock Synthetic biochemistry: the bio-inspired cell-free approach to
  commodity chemical production.
\newblock {\em Trends in biotechnology\/}, {\bf 38}(7), 766--778.

\bibitem[Durant {\em et~al.}(2002)Durant, Leland, Henry, and
  Nourse]{durant2002reoptimization}
Durant, J.~L., Leland, B.~A., Henry, D.~R., and Nourse, J.~G. (2002).
\newblock Reoptimization of mdl keys for use in drug discovery.
\newblock {\em Journal of chemical information and computer sciences\/}, {\bf
  42}(6), 1273--1280.

\bibitem[Gao {\em et~al.}(2018)Gao, Yang, Wu, Zhou, Lu, and
  Hu]{gao2018recommendation}
Gao, L., Yang, H., Wu, J., Zhou, C., Lu, W., and Hu, Y. (2018).
\newblock Recommendation with multi-source heterogeneous information.
\newblock In {\em IJCAI International Joint Conference on Artificial
  Intelligence\/}.

\bibitem[He {\em et~al.}(2017)He, Liao, Zhang, Nie, Hu, and Chua]{he2017neural}
He, X., Liao, L., Zhang, H., Nie, L., Hu, X., and Chua, T.-S. (2017).
\newblock Neural collaborative filtering.
\newblock In {\em Proceedings of the 26th international conference on world
  wide web\/}, pages 173--182.

\bibitem[Jiang {\em et~al.}(2021)Jiang, Liu, and Hassoun]{jiang2021learning}
Jiang, J., Liu, L.-P., and Hassoun, S. (2021).
\newblock Learning graph representations of biochemical networks and its
  application to enzymatic link prediction.
\newblock {\em Bioinformatics\/}, {\bf 37}(6), 793--799.

\bibitem[Kanehisa and Goto(2000)Kanehisa and Goto]{kanehisa2000kegg}
Kanehisa, M. and Goto, S. (2000).
\newblock Kegg: kyoto encyclopedia of genes and genomes.
\newblock {\em Nucleic acids research\/}, {\bf 28}(1), 27--30.

\bibitem[Kanehisa {\em et~al.}(2016)Kanehisa, Sato, Kawashima, Furumichi, and
  Tanabe]{kanehisa2016kegg}
Kanehisa, M., Sato, Y., Kawashima, M., Furumichi, M., and Tanabe, M. (2016).
\newblock Kegg as a reference resource for gene and protein annotation.
\newblock {\em Nucleic acids research\/}, {\bf 44}(D1), D457--D462.

\bibitem[Kingma and Ba(2015)Kingma and Ba]{kingma2015adam}
Kingma, D.~P. and Ba, J. (2015).
\newblock Adam: A method for stochastic optimization.
\newblock {\em ICLR\/}.

\bibitem[Kotera {\em et~al.}(2013)Kotera, Tabei, Yamanishi, Moriya, Tokimatsu,
  Kanehisa, and Goto]{kotera2013kcf}
Kotera, M., Tabei, Y., Yamanishi, Y., Moriya, Y., Tokimatsu, T., Kanehisa, M.,
  and Goto, S. (2013).
\newblock Kcf-s: Kegg chemical function and substructure for improved
  interpretability and prediction in chemical bioinformatics.
\newblock {\em BMC systems biology\/}, {\bf 7}(6), 1--17.

\bibitem[Lim {\em et~al.}(2016)Lim, Poleksic, Yao, Tong, He, Zhuang, Meng, and
  Xie]{lim2016large}
Lim, H., Poleksic, A., Yao, Y., Tong, H., He, D., Zhuang, L., Meng, P., and
  Xie, L. (2016).
\newblock Large-scale off-target identification using fast and accurate dual
  regularized one-class collaborative filtering and its application to drug
  repurposing.
\newblock {\em PLoS computational biology\/}, {\bf 12}(10), e1005135.

\bibitem[Liu {\em et~al.}(2019)Liu, Liang, and Gitter]{liu2019loss}
Liu, S., Liang, Y., and Gitter, A. (2019).
\newblock Loss-balanced task weighting to reduce negative transfer in
  multi-task learning.
\newblock In {\em Proceedings of the AAAI Conference on Artificial
  Intelligence\/}, volume~33, pages 9977--9978.

\bibitem[Liu {\em et~al.}(2016)Liu, Wu, Miao, Zhao, and
  Li]{liu2016neighborhood}
Liu, Y., Wu, M., Miao, C., Zhao, P., and Li, X.-L. (2016).
\newblock Neighborhood regularized logistic matrix factorization for
  drug-target interaction prediction.
\newblock {\em PLoS computational biology\/}, {\bf 12}(2), e1004760.

\bibitem[Liu {\em et~al.}(2021)Liu, Ma, Ouyang, and Xiong]{liu2021contrastive}
Liu, Z., Ma, Y., Ouyang, Y., and Xiong, Z. (2021).
\newblock Contrastive learning for recommender system.
\newblock {\em arXiv preprint arXiv:2101.01317\/}.

\bibitem[Mellor {\em et~al.}(2016)Mellor, Grigoras, Carbonell, and
  Faulon]{mellor2016semisupervised}
Mellor, J., Grigoras, I., Carbonell, P., and Faulon, J.-L. (2016).
\newblock Semisupervised gaussian process for automated enzyme search.
\newblock {\em ACS synthetic biology\/}, {\bf 5}(6), 518--528.

\bibitem[Mnih and Salakhutdinov(2008)Mnih and
  Salakhutdinov]{mnih2008probabilistic}
Mnih, A. and Salakhutdinov, R.~R. (2008).
\newblock Probabilistic matrix factorization.
\newblock In {\em Advances in neural information processing systems\/}, pages
  1257--1264.

\bibitem[Porokhin {\em et~al.}(2021)Porokhin, Amin, Nicks, Gopinarayanan, Nair,
  and Hassoun]{Porokhin2021Analysis}
Porokhin, V., Amin, S.~A., Nicks, T.~B., Gopinarayanan, V.~E., Nair, N.~U., and
  Hassoun, S. (2021).
\newblock Analysis of metabolic network disruption in engineered microbial
  hosts due to enzyme promiscuity.
\newblock {\em Metabolic Engineering Communications\/}, {\bf 12}, e00170.

\bibitem[Ridder and Wagener(2008)Ridder and Wagener]{ridder2008sygma}
Ridder, L. and Wagener, M. (2008).
\newblock Sygma: combining expert knowledge and empirical scoring in the
  prediction of metabolites.
\newblock {\em ChemMedChem: Chemistry Enabling Drug Discovery\/}, {\bf 3}(5),
  821--832.

\bibitem[Romero and Arnold(2009)Romero and Arnold]{romero2009exploring}
Romero, P.~A. and Arnold, F.~H. (2009).
\newblock Exploring protein fitness landscapes by directed evolution.
\newblock {\em Nature reviews Molecular cell biology\/}, {\bf 10}(12),
  866--876.

\bibitem[Sun {\em et~al.}(2019)Sun, Guo, Yang, Fang, Guo, Zhang, and
  Burke]{sun2019research}
Sun, Z., Guo, Q., Yang, J., Fang, H., Guo, G., Zhang, J., and Burke, R. (2019).
\newblock Research commentary on recommendations with side information: A
  survey and research directions.
\newblock {\em Electronic Commerce Research and Applications\/}, {\bf 37},
  100879.

\bibitem[Tawfik and S(2010)Tawfik and S]{tawfik2010enzyme}
Tawfik, O.~K. and S, D. (2010).
\newblock Enzyme promiscuity: a mechanistic and evolutionary perspective.
\newblock {\em Annual review of biochemistry\/}, {\bf 79}, 471--505.

\bibitem[Tyzack and Kirchmair(2019)Tyzack and
  Kirchmair]{tyzack2019computational}
Tyzack, J.~D. and Kirchmair, J. (2019).
\newblock Computational methods and tools to predict cytochrome p450 metabolism
  for drug discovery.
\newblock {\em Chemical biology \& drug design\/}, {\bf 93}(4), 377--386.

\bibitem[Van~der Maaten and Hinton(2008)Van~der Maaten and
  Hinton]{van2008visualizing}
Van~der Maaten, L. and Hinton, G. (2008).
\newblock Visualizing data using t-sne.
\newblock {\em Journal of machine learning research\/}, {\bf 9}(11).

\bibitem[Visani {\em et~al.}(2020)Visani, Hughes, and
  Hassoun]{visani2020enzyme}
Visani, G.~M., Hughes, M.~C., and Hassoun, S. (2020).
\newblock Enzyme promiscuity prediction using hierarchy-informed multi-label
  classification.
\newblock {\em arXiv preprint arXiv:2002.07327\/}.

\bibitem[Wang {\em et~al.}(2019)Wang, He, Wang, Feng, and Chua]{wang2019neural}
Wang, X., He, X., Wang, M., Feng, F., and Chua, T.-S. (2019).
\newblock Neural graph collaborative filtering.
\newblock In {\em Proceedings of the 42nd international ACM SIGIR conference on
  Research and development in Information Retrieval\/}, pages 165--174.

\bibitem[Wang {\em et~al.}(2021)Wang, Dong, Li, and Zhang]{wang2021multitask}
Wang, Y., Dong, L., Li, Y., and Zhang, H. (2021).
\newblock Multitask feature learning approach for knowledge graph enhanced
  recommendations with ripplenet.
\newblock {\em Plos one\/}, {\bf 16}(5), e0251162.

\bibitem[Webb {\em et~al.}(1992)Webb {\em et~al.}]{webb1992enzyme}
Webb, E.~C. {\em et~al.} (1992).
\newblock {\em Enzyme nomenclature 1992. Recommendations of the Nomenclature
  Committee of the International Union of Biochemistry and Molecular Biology on
  the Nomenclature and Classification of Enzymes.}
\newblock Number Ed. 6 in 6. Academic Press.

\bibitem[Weinberger and Saul(2009)Weinberger and Saul]{weinberger2009distance}
Weinberger, K.~Q. and Saul, L.~K. (2009).
\newblock Distance metric learning for large margin nearest neighbor
  classification.
\newblock {\em Journal of machine learning research\/}, {\bf 10}(2).

\bibitem[Xue {\em et~al.}(2017)Xue, Dai, Zhang, Huang, and Chen]{xue2017deep}
Xue, H.-J., Dai, X., Zhang, J., Huang, S., and Chen, J. (2017).
\newblock Deep matrix factorization models for recommender systems.
\newblock In {\em IJCAI\/}, volume~17, pages 3203--3209. Melbourne, Australia.

\bibitem[Zhang {\em et~al.}(2020)Zhang, Zhang, Wu, Ou-Yang, Zhao, and
  Li]{zhang2020graph}
Zhang, Z.-C., Zhang, X.-F., Wu, M., Ou-Yang, L., Zhao, X.-M., and Li, X.-L.
  (2020).
\newblock A graph regularized generalized matrix factorization model for
  predicting links in biomedical bipartite networks.
\newblock {\em Bioinformatics\/}, {\bf 36}(11), 3474--3481.

\bibitem[Zheng {\em et~al.}(2013)Zheng, Ding, Mamitsuka, and
  Zhu]{zheng2013collaborative}
Zheng, X., Ding, H., Mamitsuka, H., and Zhu, S. (2013).
\newblock Collaborative matrix factorization with multiple similarities for
  predicting drug-target interactions.
\newblock In {\em Proceedings of the 19th ACM SIGKDD international conference
  on Knowledge discovery and data mining\/}, pages 1025--1033.

\bibitem[Zhu {\em et~al.}(2017)Zhu, Zhang, He, Wu, Zhou, Zhang, and
  Yu]{zhu2017broad}
Zhu, J., Zhang, J., He, L., Wu, Q., Zhou, B., Zhang, C., and Yu, P.~S. (2017).
\newblock Broad learning based multi-source collaborative recommendation.
\newblock In {\em Proceedings of the 2017 ACM on Conference on Information and
  Knowledge Management\/}, pages 1409--1418.

\end{thebibliography}

\end{document}